\documentclass[sn-mathphys,Numbered]{sn-jnl}

\usepackage{subcaption}
\usepackage{graphicx}%
\usepackage{multirow}%
\usepackage{amsmath,amssymb,amsfonts}%
\usepackage{amsthm}%
\usepackage{mathrsfs}%
\usepackage[title]{appendix}%
\usepackage{xcolor}%
\usepackage{textcomp}%
\usepackage{manyfoot}%
\usepackage{booktabs}%
\usepackage{algorithm}%
\usepackage{algorithmicx}%
\usepackage{algpseudocode}%
\usepackage{listings}%
\usepackage{todonotes}

\begin{document}

\title[{On-policy Actor-Critic Reinforcement Learning for Multi-UAV Exploration} ]{{On-policy Actor-Critic Reinforcement Learning for Multi-UAV Exploration} }


\author[1]{ \sur{Ali Moltajaei Farid}}\email{ali.farid@kntu.ac.ir}

\author[1]{ \sur{Jafar Roshanian}}\email{roshanian@kntu.ac.ir}

\author*[2]{ \sur{Malek Mouhoub}}\email{mouhoubm@uregina.ca}

\affil[1]{\orgdiv{Intelligent Control Systems Institute}, \orgname{K. N. Toosi University of Technology}, \orgaddress{\street{Daneshgah Street}, \city{Tehran}, \postcode{1656983911}, \country{Iran}}}

\affil[2]{\orgdiv{Department of Computer Science}, \orgname{University of Regina}, \orgaddress{\street{Wascana Pkwy}, \city{Regina}, \postcode{S4S0A2}, \country{Canada}}}

 \setlength {\marginparwidth }{2cm}

\abstract{

Unmanned aerial vehicles (UAVs) have become increasingly
popular in various fields, including precision agriculture,
search and rescue, and remote sensing. However, exploring
unknown environments remains a significant challenge. This
study aims to address this challenge by utilizing on-policy
Reinforcement Learning (RL) with Proximal Policy Optimization
(PPO) to explore the  {two dimensional} area of interest with multiple UAVs.  The UAVs will avoid collision with obstacles and each other and do the exploration in a distributed manner.  The
proposed solution includes actor-critic networks using deep
convolutional neural networks  {(CNN)} and long short-term memory
(LSTM) for identifying the UAVs and areas that have already been covered.
Compared to other RL techniques, such as  policy gradient (PG) and asynchronous advantage actor-critic (A3C), the simulation results demonstrate the superiority of the proposed
PPO approach.  Also, the results show that combining LSTM with CNN in critic can improve exploration. Since the proposed exploration has to work in unknown environments, the results showed that the proposed setup can complete the coverage when we have new maps that differ from the trained maps. Finally, we showed how tuning hyper-parameters may affect the overall performance.  

}




\maketitle

\section{Introduction}

In recent years, the technology of unmanned aircraft systems has evolved rapidly. In particular, unmanned aerial vehicles (UAVs) today reach inaccessible and dangerous places without needing humans to be on-site. Also thanks to recent advances in autonomous navigation and artificial intelligence (AI) coordinated fleets of UAVs. UAVs are particularly useful for time-sensitive operations, such as the delivery of medical equipment or environmental reconnaissance for search and rescue, but despite the potential benefits of implementing a fleet of UAVs, properly designed coordination strategies are required to exploit them in the best possible way \cite{almahamid2022autonomous}. {In addition, UAVs have limited access to computational resources, which creates a significant need for effective computational algorithms for on-board processors. We need efficient algorithms in terms of performance and computations, which remains a challenge.}

{Single-agent systems can complete their tasks, but they are often slow and suffer from limitations such as limited energy resources. On the other hand, multi-agent systems are more time-efficient. There has been a lot of research in this field, particularly in the navigation of multi-agent systems like UAVs, Unmanned Ground Vehicles, and Underwater Autonomous Vehicles. Multi-agent planning and coordination are two similar terms that belong to multi-agent systems. Multi-agent planning involves determining the sequence of feasible actions of the agents to achieve individual goals while maintaining optimality. Coordination, on the other hand, refers to the effective interaction among the agents to serve the purpose of all the agents.}

{It is crucial to note that the exploration problem's type varies depending on the case. For instance, certain problems can be assumed to be two-dimensional, in other words, UAVs fly at a constant height from the terrain. Additionally, some problems may require consideration in three-dimensional space due to the terrain's hills. However, for simplicity, this paper will focus on a two-dimensional space problem.
} 

{It is crucial to consider the type of UAVs used in a study. Multi-rotor, fixed-wing, or hybrid UAVs that combine the features of fixed-wing and multi-rotors are commonly used. Multi-rotors have higher agility but are limited in terms of payload capacity, while fixed-wing UAVs can carry more payload but have limitations when it comes to quick turns. Fixed-wing UAVs require additional computations such as Dubins curves to handle safe turning. It is important to note that the dynamics of the UAVs are different at a high level. For this study, only multi-rotor UAVs are considered. We are not taking into account the dynamical equations of motion of the UAV, but rather assume it as a point that can pass through waypoints.
}

{After identifying the area of interest, we have various techniques to break down the area into smaller subareas to assign each UAV. Each subarea can be divided into a number of cells. Subsequently, we need to figure out a way to navigate from these cells while considering constraints such as avoiding obstacles, optimizing energy consumption, and other similar factors. 
}

{Reinforcement learning has been used in mapping in the literature extensively. } For known environments, Chen et al. \cite{chen2019adaptive} proposed the adaptive deep path (ADPath) based on cellular decomposition followed by deep reinforcement learning. The authors first use  {boustrophedon cellular decomposition} to divide the problem into multiple cells and then use a deep Q learning network (DQN) to find the order of covering the cells. Similarly, Piardi et al. \cite{piardi2019coverage} use DQN to obtain an optimal path where {each cell} is visited only once. After analyzing the {cell-based} environment, Kiaw et al. \cite{kyaw2020coverage} have used RL with recurrent neural networks to solve the traveling salesperson problem (TSP).

In \cite{theile2020uav}, a double DQN is used to learn a discrete control policy for a UAV, which has been used to balance the use of energy from limited power and coverage. For unknown environments, in \cite{heydari2021reinforcement} Heydari et al. formulated {coverage path planning} as an optimal stopping time problem, where the reward function and the state representation can be defined naturally. Niroi et al. \cite{niroui2019deep} present a boundary-based method for {coverage path planning}. The authors use RL to calculate the cost of each boundary node and then apply the A* algorithm to move to the selected node. In \cite{saha2021deep}, a variant of DQN with prioritized experience replay is used to cover new unseen environments with minimal additional sample complexity effectively. Then, the advantage actor-critic method is used for coverage with minimal overlap in unknown indoor environments with dynamic obstacles \cite{saha2021online}.

Hu et al. \cite{hu2020voronoi} also applied RL in a continuous environment. They consider multi-robot exploration and use Voronoi partitioning to divide robots into free space. The authors use a frontier approach to calculate new waypoints for each robot in an action planner. The robots move to their designated path. The robots navigate to their designated point using a low-level RL agent, which receives the relative position of the waypoint along with lidar data as input.

There are non-RL approaches for exploration such as evolutionary based approaches, While heuristic approaches such as evolutionary algorithms have been used in the literature \cite{farid2022evolutionary, farid2023multi}, they are computationally hungry and challenging to implement in online embedded systems. In contrast, RL approaches are computationally hungry in the training phase only and can be seen as an option in online embedded systems \cite{hodge2021deep}. 

Our proposed approach orchestrates the coverage of two-dimensional environments with a flexible number of UAVs. However, the experiments are restricted to a fleet size ranging from 3 to 8 UAVs. This design holds potential for application in systems dedicated to exploring or mapping unknown environments with a multitude of UAVs. There is no such specific RL approach to the best of the author’s knowledge. Such an approach is essential with strategies based on RL applied to multi-agent systems. The mapping is done by encouraging the agents to learn a competitive behavior in which each UAV tries to maximize its explored areas. This competitive setup, where all UAVs have the same objective,
enables full mapping capabilities under test conditions.  Tables 1 and 2 compare the main features of the proposed approach vs the literature.

The rest of the paper is as follows: First, we briefly
present RL. Next, our simulator specifications will be described. In Section 4, we demonstrate our
simulation results and comparisons. Finally, in Section 5 we conclude our
paper by investigating the future directions.

\begin{table}[t!]  
\caption{ {The literature vs our proposed approach}}
\label{table_example}
\begin{tabular}{|l|l|l|l|}
\hline
Approach & Homogeneous  & Heterogeneous & Vehicle
\\
\hline
Multi-objective optimization \cite{farid2023multi} & Y & Y & MR, FW 

\\
\hline
Single-objective optimization \cite{farid2022evolutionary} & Y & N & MR

\\
\hline
The proposed approach& Y & N & MR

\\
\hline

\end{tabular}
\end{table}

\begin{table}[t!]  
\label{t1}
\caption{{The pros and cons of the literature and our proposed approach}}
\label{table_example}
\begin{tabular}{
|p{0.25\linewidth}|p{0.25\linewidth}| p{0.25\linewidth}|}
\hline
Approach &  Advantages & Drawbacks
\\
\hline
Multi-objective optimization \cite{farid2023multi} & •	Can control UAV as either offline or online controller

 & •	Time-consuming for vast environment planning
 
•	High computation

•	Needs proper tuning of parameters

\\
\hline
Single-objective optimization \cite{farid2022evolutionary} & •Less computation comparing to Multi-objective optimization

•Can control UAV as either offline or online controller

 & •	less quality comparing to Multi-objective optimization
 
•	Needs proper tuning of parameters

\\
\hline
The proposed approach
&
• Small amount of computation for execution phase

• Can modify the actions in case of exploring a new environment (different from the trained environment)

 & 
 • Too high computation for training phase

•	Needs proper tuning of parameters

\\
\hline
\end{tabular}
\end{table}

\section{ Reinforcement Learning (RL)}

 In RL, the main elements in defining a given problem are as follows: agents, environment, actions, rewards, and observation. The agent is an entity that interacts with the environment. This agent can take actions that determine changes in the environment. The agent makes observations about the state of the environment and receives rewards from the environment based on the action performed at a particular moment. The environment is everything that is outside the agent. Next, actions taken by the agent can be either discrete or continuous. The agent can perform a set of enumerated and unique actions in a discrete space. For example, an agent that can only move in 4 directions, such as in a video game. In a continuous space, the actions can have a real value in a specific range; for example, the agent can move in any direction specified by a real number. A scalar value is the reward for an interaction between the agent and the environment. This reward is seen as an agent decision quality to maximize as much as possible. Finally, with observation, the agent can distinguish between different states.

RL algorithms can be classified into off-policy methods and on-policy methods. Off-policy algorithms evaluate and improve a target policy different from those used to explore the environment and generate experiences. The most common methods in this category are the Q-learning family (DQN, Double DQN), the actor-critic methods such as Soft Actor Critic (SAC), PG, deep deterministic policy gradient (DDPG), and twin-delayed deep deterministic policy gradient (TD3), which often collect a large set of state transitions (i.e., experiences) \cite{azar2021drone}. At update time, off-policy methods randomly sample a batch of transitions to improve their policy and often have heuristic strategies such as the Greedy method.
On the other hand, on-policy methods directly improve a target policy on top of the current policy. Common methods in this category include on-policy time difference learning such as SARSA, critical on-policy methods AC, Advantage Actor Critic (A2C),  A3C, and in-field policy slope methods with (REINFORCE) \cite{azar2021drone}. Also, trust region policy optimization (TRPO), PPO, and others use long-term cumulative rewards instead of learning time differences to optimize long routes \cite{guan2023cooperative}.

{Multi-agent reinforcement learning, or MARL, is a broad research area where agents cooperate and compete.In MARL, agents learn their own policies in a shared environment, where the environment are influenced with all agents. One way of learning agents from the environment is decentralized way of learning treating other agents as part of the environment. In other approach agents can learn a joint policy.  The above approaches might lead to non-stationary stability. To address this issue recently most researcher, rely on centralized training and decentralized execution approach, where agent can learn with global access to other agents and environment during the training and local information during the execution. Exploration in RL is a challenge as multi-agents should choose the best action to maximize rewards or acquire more information from the environment.}

Among the applications of MARL,  efforts for compelling gameplay in popular video games can be mentioned \cite{laurent2021flatland}. Coordinated navigation for multiple UAVs is another research direction \cite{frattolillo2023scalable}. The navigation and exploration tasks of multi-agent systems have been investigated in the literature using several techniques, including nature-inspired methods \cite{cui2022survey} and RL-based methods. Some approaches use pre-processing or clustering techniques about RL agents \cite{zhang2021multi}. In contrast, others apply specific MARL algorithms such as multi-agent deep deterministic policy gradient (MADDPG) \cite{du2021cooperative} or multi-approximate Q-learning \cite{zhang2021multi}.

There are several advantages of using  {multi-agent} instead of single-agent using RL: First, when multiple agents solve the same problem, agents can share their experiences, leading to faster and better problem-solving. Examples of how agents can share experience are information exchange using communication, or a less skilled agent can use a more skilled agent as a teacher. The second advantage is about using multiple agents to solve a decentralized problem. In this case, agents can work in parallel, which speeds up problem-solving. Thirdly, when one of the agents is lost for any reason, other agents can take over and divide the tasks among themselves. For this reason, using multiple agents makes problem-solving more robust compared to using a single agent. In addition,  MARL can quickly scale to larger problems.

On the other hand,  the use of multi-agent also comes with several challenges: As already explained above, a fundamental problem is the large amount of state-action combinations, making estimation difficult with increased agents. The second challenge is how to define the learning goal. In particular, returns are correlated to various agents, which are impossible to maximize independently. This challenge leads to a trade-off between stability and adaptability. Stability refers to the convergence of the algorithm to a fixed policy. At the same time, adaptability ensures that an individual agent's policy is adjusted when other agents' policies change. Consistency is desirable because it facilitates analysis and provides meaningful performance guarantees. Also, when one agent is stable, it reduces noise for other agents and makes adaptation easier.
On the other hand, consistency is desirable because the behavior of other agents is vastly unpredictable. The lack of stationarity in the MARL algorithm relates to the above issue as agents adapt to the policy of other agents. However, these other agents, in turn, readjust their policy based on the adjustment. Therefore, the best policy for one agent moves with the policy of other agents. Also, in MARL, several factors make exploration and exploitation more complicated. Agents explore not only to gather more information about the environment but also about the behavior of other agents. However, if one agent explores too much, this destabilizes the other agent, making learning more difficult for the current agent.  

In the proposed system, we applied several UAVs for exploration. As the number of UAVs (agents) increases, the time for training increases dramatically, which requires a high amount of processing power. The system's stability was preserved by properly defining rewards and setting a proper amount for hyper parameters, described in the following sections. We fine-tuned the entropy loss weight to regulate the balance between exploration and exploitation in this configuration. We specify a scalar value between 0 and 1 for the entropy loss weight. A higher value encourages agent exploration by penalizing certainty in action selection \cite{schulman2015trust}.

\section{System Design}
\begin{figure}[t!]
\centering
\includegraphics[width=.7\textwidth]{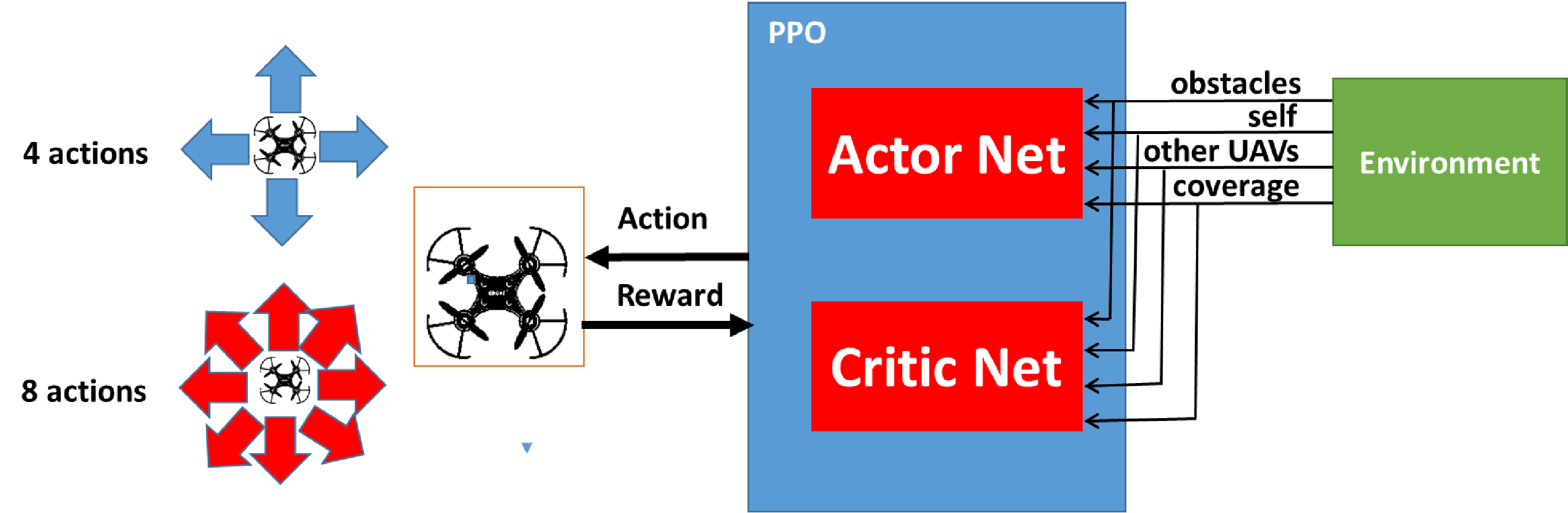}%
\caption{PPO actor-critic setup with different action spaces}
\label{Fig01}
\end{figure}

It is very costly to train the RL system for real-world experiments, so similar to other literature; we use a simulator to simulate the exploration mission. The environment is a grid type with several cells in which each UAV can stay inside a specific cell at a specific time. The cell's dimensions are based on the assumed UAV camera field of view (FOV) at a specific height. For simplicity, we assumed that the UAVs fly at a certain height during the whole mission. In addition, we simulate the non-flying-zone (NFZ) areas with obstacles where UAVs will be penalized in case they collide with these areas. For training the RL, we used the simulator where the states are derived for training, and actions are chosen based on the states, as shown in Figure 1. In the following, we will first formulate the problem. Then, we review the RL method that we have used (Proximal Policy Optimization (PPO) reinforcement learning ). Next, we will discuss the observation, actions,  rewards, and actor-critic networks.

\subsection{Problem formulation}
{Let us assume a set of  UAVs G=$\{g_1, g_2, g_3,\ldots,g_i\ldots\}$, whose $g_i$ denotes UAV agent $i$. Both the initial position and initial numbers of set G are known. The positions of agents $gq_i$ at time $t$, are shown with  $p_i(t)$. Each $ga_i$ has a limited FOV and a limited camera range ($C_R$). Each UAV $g_i$ has a downward camera, which maps a bounded size of $n$x$m$ in the specific height.}
In this simulation, we did not consider the dynamic specification of the UAV. We assume the UAV to be a single point and only focus on its path planning during the exploration.

This design aims to find a set of waypoints to meet the condition, which is improving the efficiency of exploration time by collaboration among UAVs. The reinforcement learning algorithm is used to learn how to use UAVs and their routing to explore the environment. In the reinforcement learning method, each agent is trained in a centralized manner during training; in other words, all UAVs have complete information from the training environment; this information includes the location of all UAVs as well as the already explored areas, all the obstacles, and cells that are not permitted to fly over. Next, each UAV will decide in a decentralized manner during and after the training phase. 

 We developed a simulator where the user can define the environment, including the dimensions, the NFZ cells, and obstacles. In addition, the user can define the number of UAVs that fulfill the mission. Besides, the user can define the training phase settings such as the maximum number of episodes, the type of RL (PPO, PG, or A3C), the RL's training hyperparameters, the architecture of actor and critic deep convolutional networks, whether as CNN or LSTM. Also, the type of training can be either centralized or decentralized. In decentralized training, each agent gathers its own set of experiences during the episodes and learns independently from those experiences.
On the other hand, in centralized training, the agents share the collected experiences and learn from them collectively. After the completion of a trajectory, the actor and critic functions are synchronized between the agents. When the user sets these settings, the simulator can train the system with the environment and compare the performance in new environments. The simulator can find the number of collisions, the energy consumed, and the total time of the mission.

 \subsection{Multi-Agent Markov Decision Process}
The proposed multi-agent system can be modeled as a multi-agent Markov Decision Process (MDP). The MDP is defined as $(S, A, T, R, Z, O, n, \gamma)$, where $S$ and $A$ denote each agent's states and actions, respectively. At each time step $t$, each agent's target policy $\zeta_{i_t}$, where $i \in {1, . . . , n}$, selects an action $a_{t_i} \in A$. All selected actions form a joint action $a_t \in A^n$. The transition function $T$ maps the current state $s_t$ and the joint action to a distribution over the next state $s_{t+1}$, i.e., $T: S	\times A^n 	\to	\Delta(S)$. All agents receive a collective reward $r^t \in R$ according to the reward function $R: S 	\times A^n 	\to R$. The objective of all agents' policies is to maximize the collective return $\sum_t \gamma^t r^t$, where $\gamma \in [0, 1]$ is the discount factor, and $r^t$ is the collective reward obtained at time step $t$. Each agent $i$ observes a local observation $o_{t_i} \in Z$ according to the observation function $O: S \to Z$. We model the problem as an MDP and solve it using the Proximal Policy Optimisation (PPO) method as the following subsection. Since in recent papers \cite{theile2023learning,zhan2022multiple}, PPO shows a good performance, PPO is chosen as an RL solver of the exploration. Since PPO is on-policy RL, we compared two other on-policy solvers with the PPO.

\subsection{Proximal Policy Optimization (PPO)}

PPO is a family of model-free reinforcement learning algorithms developed at OpenAI in 2017. PPO is a policy-based method that uses the same policy for exploration and exploitation. 
PPO is an expanded and simplified version of TRPO and has been used in the literature. PPO learns very slowly and needs more time than off-policy methods. However, against unknown things, it can be learned online, which is very important. 
PPO iteratively updates the policy by sampling paths from the environment and computing the policy gradient, which is the direction that improves the expected return. However, unlike other policy gradient methods, PPO does not use a constant learning rate or confidence region to control the step size. Instead, PPO uses a truncated objective function that penalizes significant policy changes. This way, PPO prevents the policy from overfitting or collapsing into a suboptimal solution. We define the relationship between new and old policies as:

\begin{equation} \label{Eq_ej}
r(\theta_1)=\frac{\pi_{\theta_1}(a|s)}{\pi_{\theta_1^{old}}(a|s)}
\end{equation}

then the new actor function in PPO is as follows:

\begin{equation} \label{Eq_ej1}
\mathop{\mathbb{E}} [A(s)r(\theta_1)-\beta S(\pi_{\theta_1}|s)]
\end{equation}

if $\hat{r}(\theta_1)=clip(r(\theta_1),1-\epsilon,1+\epsilon)$ then the actor function will be:

\begin{equation} \label{Eq_ej2}
J^{PPO}(\theta_1)=\mathbb{E}[min(A(s).r(\theta_1), A(s).\hat{r}(\theta_1))-\beta S(\pi_{\theta_1} |s)]
\end{equation}
where $(\pi_{\theta_1|s})$ is the entropy of policy $\pi_{{\theta}_1}$ given state $s$, and $\beta$ is an entropy regularization constant. $\epsilon$ is clipping Hyperparameter. $A$  {represents an action in state} $s$. In PPO the Critic uses the same clipping technique used by the Actor, but instead of keeping the minimum between the clipped and the non-clipped objective, it keeps the maximum. 

The discrete type of action is chosen to reduce the complexity of the simulation. In this way, each UAV will have four, five, or eight actions that help the UAV to move in the desired direction. Also, several reward functions have been compared. The obtained output shows that this type of design is more suitable for foraging and mapping systems. In the proposed method, instead of sending the information of each UAV, the deep convolutional network method is used, where the location of each UAV, obstacles, and previously searched paths are extracted from the image of each simulation moment. In this way, each UAV will access the path previously covered by the UAVs and remove the obstacles. By using the learned policy, it will be able to make the appropriate decision.

\subsection{Observation}
The following method is adopted to parse the map. First, the user defines the desired space for exploration. The space is then broken down into a set of cells. Therefore, the environment map is a matrix with a value assigned to each cell representing its current state. Each cell is the FOV area or field of view a UAV covers at a given altitude. We assume that the user specifies this height, which is the same for all UAVs. For simplicity, we assume that all UAVs are identical and have the same FOV. Therefore, in a specific time instance $t$, each UAV occupies one of the cells. We assume the UAVs have the same constant speed, so each agent can only travel to the next immediate cell from time $t$ to $t + 1$. In order to preserve the characteristics of the environment spaces based on its current state, i.e., agents, obstacles, and explored areas, we assign a different color, dividing it into four channels, including obstacle, self, other UAVs, and coverage channels.

\subsection{Action space}
 The design of the action space affects the strategy space for task completion. A complex action design makes the strategy space of the problem too significant and makes it difficult to train the reinforcement learning algorithm to converge. On the other hand, a too-simple action design makes the strategy space too small, and it might need to find a suitable strategy to perform the task, or it is less effective. Therefore, action design in the first part includes a small number of discrete actions, a total of 4, 5, or 8 actions. For example, when we have five actions, they include going up, going down, staying steady, turning left, and turning right. The action space for 4 and 8 actions is depicted in Figure 1. 

\subsection{Energy}
 The path of the UAV can be considered as a movement. When the UAV moves in a straight line at a uniform speed in a sequence of actions, it flies for $t$ seconds, $t$ is the time interval between the two actions. Note that if there is a rotation in a sequence of actions, it should be rotated 90 degrees, and their time interval is different from $t$ and has a more significant value. Therefore, the rotation will increase the UAV's energy consumption.

\subsection{Reward design}
In deep reinforcement learning, the design of rewards directly affects the direction of model training. Quantitative evaluation of UAV actions can guide the UAV to cover the work area and obtain maximum efficiency. The effect of the reward function on the output quality is undeniable. Therefore, several reward functions have been defined so that the work output can be compared:

$R_{mov}$ should give a   negative reward, every time the UAV moves, to prevent the agent from doing invalid actions and to converge the algorithm as soon as possible.

 {$R_{cla}$ should give a   negative reward, every time the UAV collides with the obstacles (-2 for each), and UAV collides with other UAVs (-5 for each).}

$R_{cov}$ A negative bonus should be given when the UAV covers the same cell that was already explored. The restriction is to prevent the UAV from repeatedly hovering in the covered area.

$R_{NFZ}$ When the UAV flies into the no-fly zone on the map, a negative reward (penalty) should be given.

$R_{Energy}$ algorithm for choosing an action is designed considering the physical model of UAVs. Changing direction instantly requires more energy in most real-world situations due to motion limitations. Therefore, we prioritized a direct move and this constraint forced the agent to create paths with the least number of rotations possible.

\begin{equation} \label{Eq_10}
R_{Energy1}=\left\{ \begin{array}{ll}
R=R+5 &{\rm if~  3 ~consecutive ~iterations}\\
R=R+3 &{\rm if~ 2 ~consecutive ~iterations} \\
\end{array}
\right.
\end{equation}
   In equation (4), if the agent moves in the same direction consecutively, we will reward the agent. It is essential to note that the agent can access its memory to know its previous actions. 

\begin{equation} \label{Eq_101}
R_{Energy2}=\left\{ \begin{array}{ll}
R=R-5 &{\rm if~   turn ~in~opposite~direction}\\
R=R &{\rm  else} \\
\end{array}
\right.
\end{equation}

When the UAVs reach the end of already set number of episodes or if the UAV covers all the cells, the following reward is considered:

\begin{equation} \label{Eq_9}
R_1=F\left(\frac {C_{cov}\cdot i_{max}}{C_{total}\cdot i_{mission}}\right)
\end{equation}

 {where} $C_{cov}$ and $C_{total}$   represent the current covered area and total number of cells.

\begin{equation} \label{Eq_91}
R_2=F
\end{equation}
where $F$ is a constant. The goals of these functions include the following:

\begin{enumerate}
\item Minimizing the penalty related to hitting an obstacle or another UAV 

\item Energy optimization: more energy will lead to a larger penalty

\item Time optimization: the speed of completing the search will increase the reward 
\end{enumerate}

These functions are used in three different types of reward functions as follows:

\begin{equation} \label{Eq_92}
T_1=R_{Energy1}+R_1+R_{cov}+R_{mov}+R_{NFZ}+R_{cla}
\end{equation}
\begin{equation} \label{Eq_93}
T_2=R_{Energy1}+R_2+R_{cov}+R_{mov}+R_{NFZ}+R_{cla}
\end{equation}
\begin{equation} \label{Eq_94}
T_3=R_{Energy2}+R_1+R_{cov}+R_{mov}+R_{NFZ}+R_{cla}
\end{equation}

\subsection{Actor-critic Networks}
 We have used two types of networks in our actor-critic configurations. First, we applied deep convolutional networks to extract the position of all UAVs, obstacles, and the paths that the team of UAVs had already searched. The configuration is shown in Figure 2. We used LSTM or long short-term memory in another configuration, as shown in Figure 3.   {In both   figures, the input is the image of the environment, while the output is the chosen action or action values in actor or critic setups, respectively. Actor critic architectures can improve the quality of training.} In this way, the UAV has access to a short memory of the previous time, while the first type does not have access to the memory. Our approach involves utilizing the PPO algorithm to train the agent and incorporating LSTM to extract time-series features for an initial state of a specific time window length. Unlike the previous deep method that employed actor deep convolutional network in agent training, we opted for LSTM as it is a recurrent neural network that can learn order dependence in sequence prediction problems. Similar to \cite{zou2022novel}, we utilized LSTM as a feature extractor in the actor. Specifically, we leveraged the memory property of LSTM to improve coverage path planning.

\begin{figure}[t!]
\centering
\includegraphics[width=.7\textwidth]{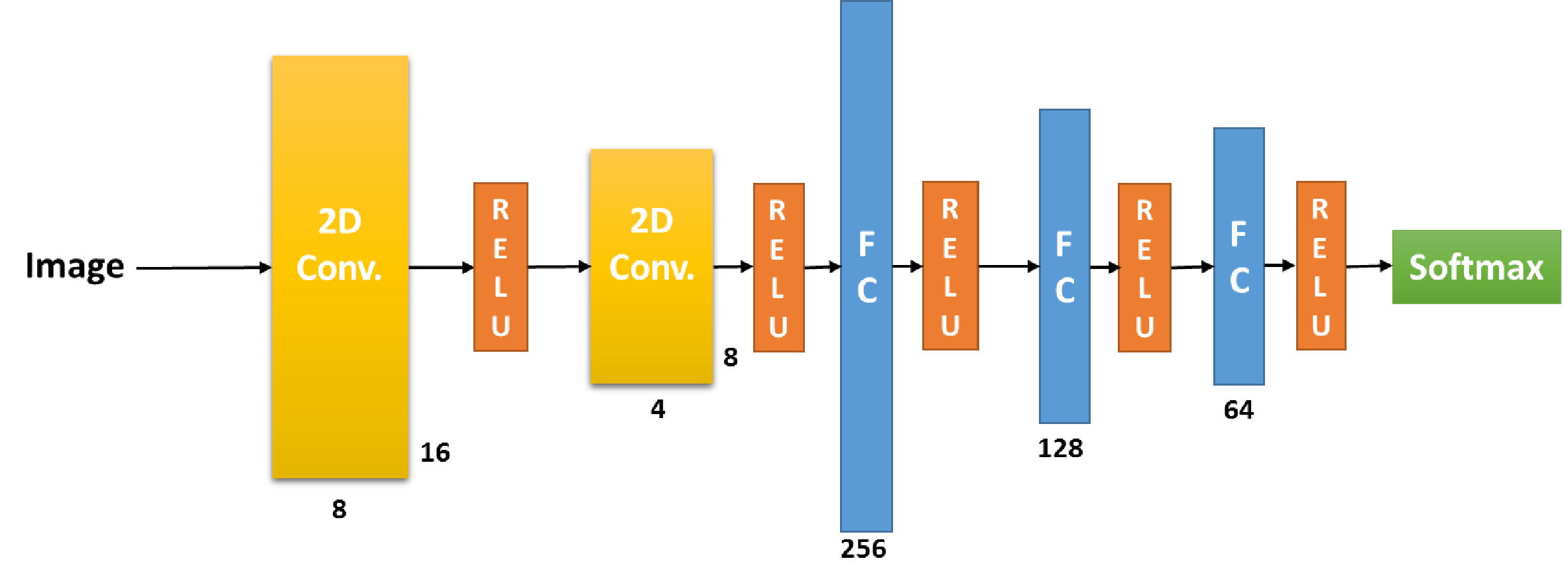}%
\caption{The actor network configuration}
\label{Fig02}
\end{figure}

\begin{figure}[t]

\centering
\subcaptionbox {}{ \includegraphics[width=0.7\textwidth]{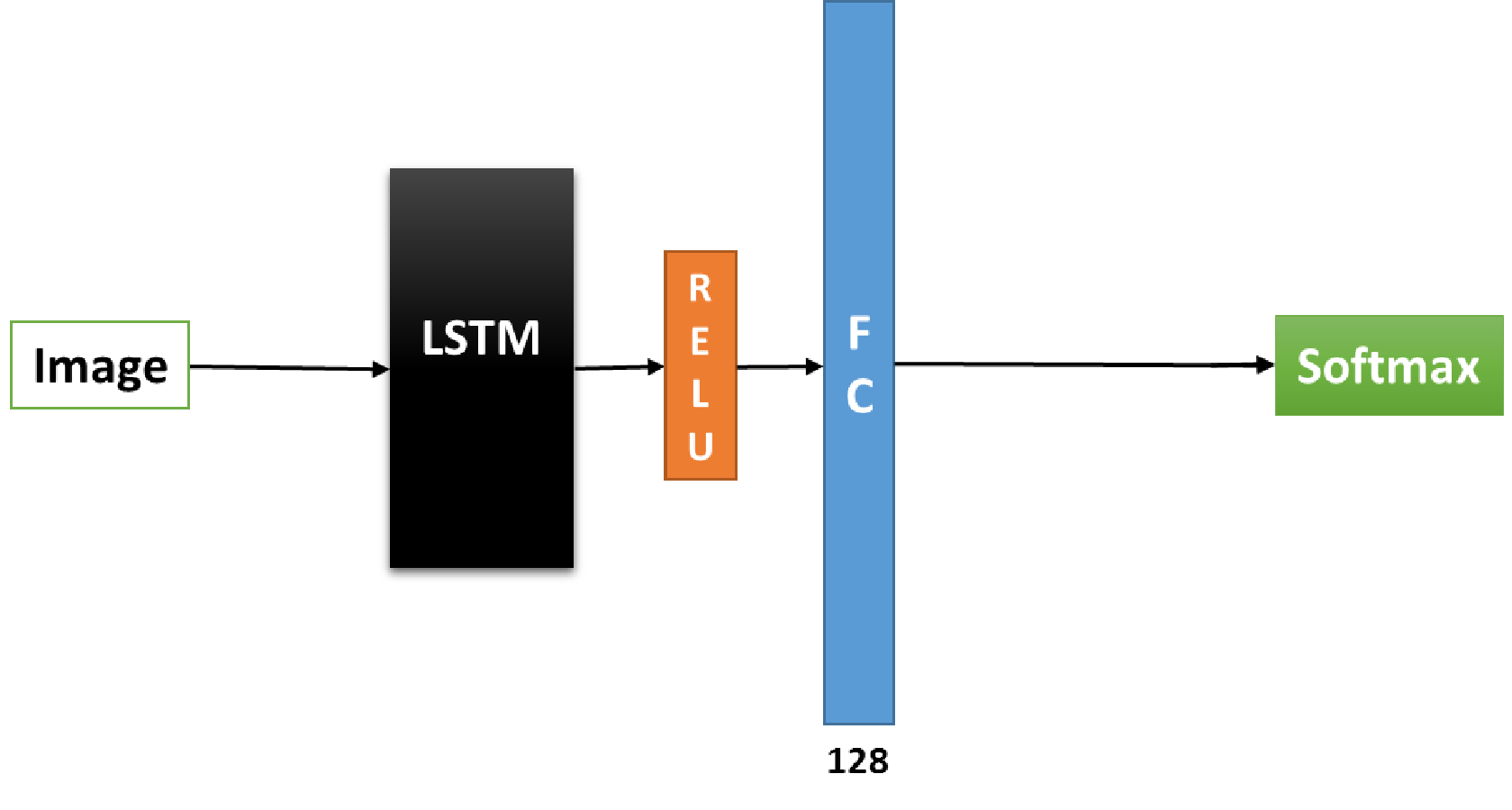}}
\\ 
\subcaptionbox {}{\includegraphics[width=0.7\textwidth]{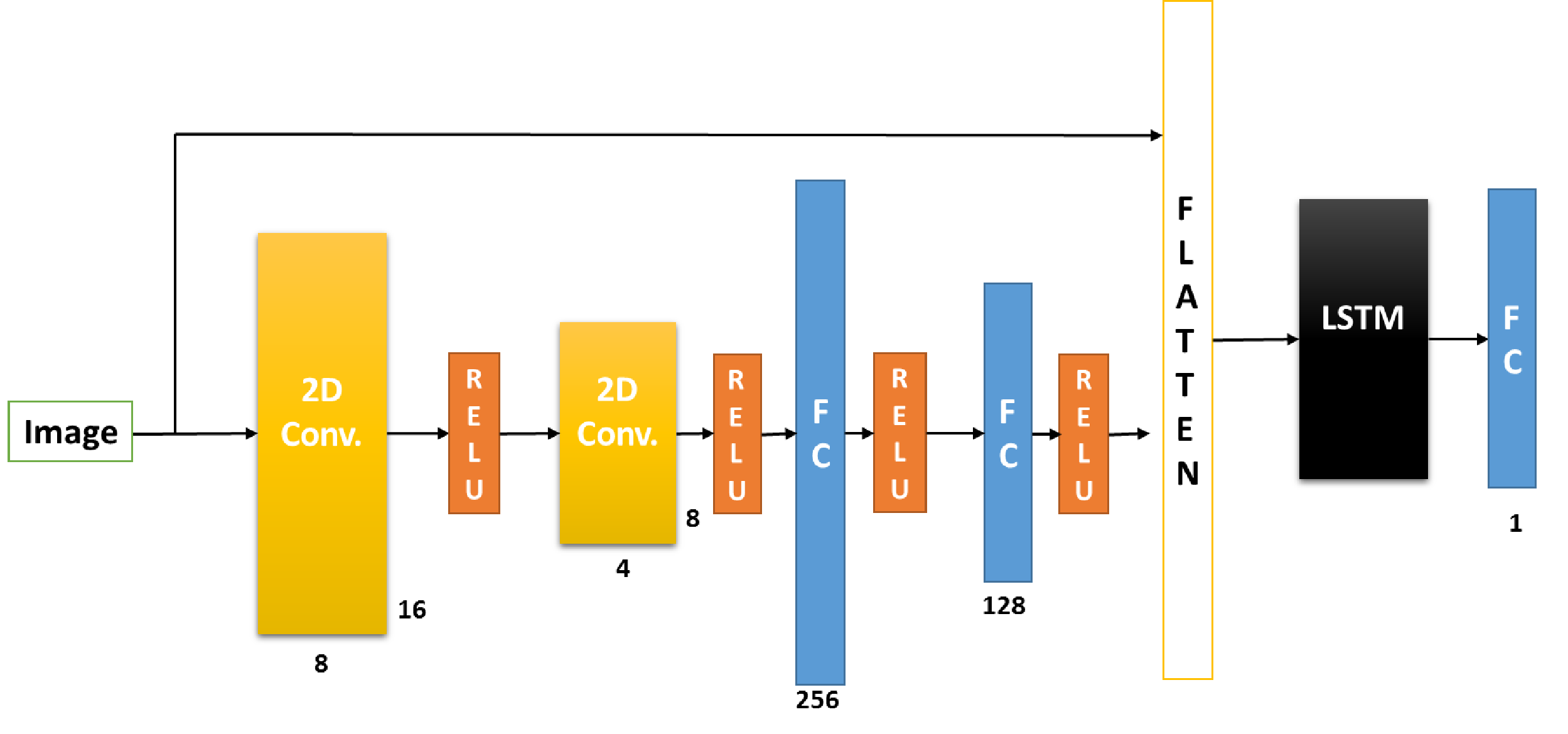}}%
\caption{(a) The actor  and (b) critic LSTM networks configuration.}
\label{fig3}
\end{figure}

\section{ Simulation Results}

 We developed our simulator using MathWorks Matlab 2022, running on a desktop PC with an Intel Core i7 processor and 32 GB RAM. According to Figure 4, we trained the RL with E1, and in some cases, we tested the performance in E2 and E3 environments. All the mentioned environments have a size of 12x12 in  a two dimensional space). 
 In Figure 4, the black cells represent the NFZ or obstacles, while other colors represent the UAVs as a circle.  {In all experiments, unless mentioned otherwise, we averaged the
performance of 3 UAVs in the specified mission, and we only show one plot.}

As shown in Figure 5, the system's efficiency increases and approaches 100 percent with the number of UAVs. However, as the complexity of the environment increases, this diagram will change, and more UAVs may be needed to increase efficiency compared to the complexity of the environment.

In Figure 6, the system's performance in preventing accidents has been checked, which shows that T2 and T3 needed a smaller number of corrections due to having a collision penalty, which shows the adaptive learning of the system.

\begin{figure}[t!]
\centering
\includegraphics[width=.7\textwidth]{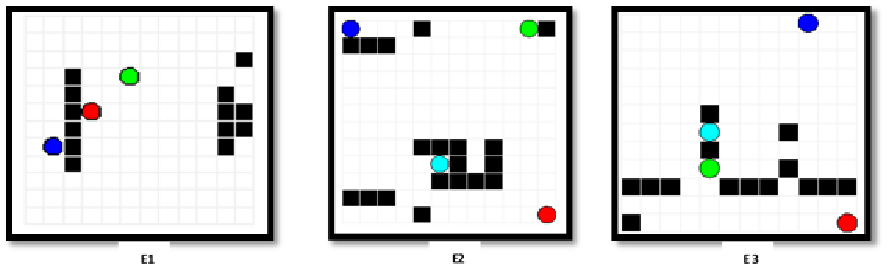}%
\caption{The E1(left), E2 (middle), and E3 (right) environments}
\label{Fig04}
\end{figure}

\begin{figure}[t!]
\centering
\includegraphics[width=.7\textwidth]{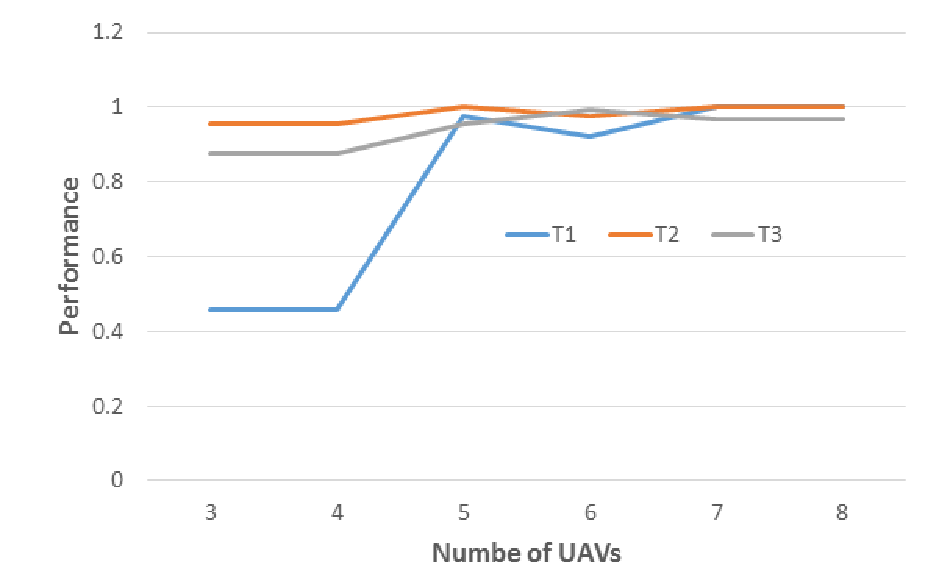}%
\caption{The exploration performance with different numbers of UAVs}
\label{Fig05}
\end{figure}

\begin{figure}[t!]
\centering
\includegraphics[width=0.7\textwidth]{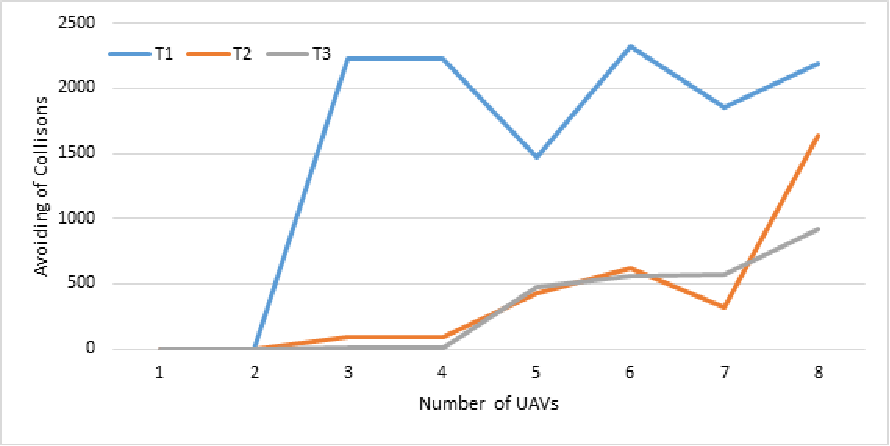}%
\caption{ {The number of UAVs versus the number of avoided collisions}}
\label{Fig06}
\end{figure}

\begin{figure}[t!]
\centering
\includegraphics[width=.7\textwidth]{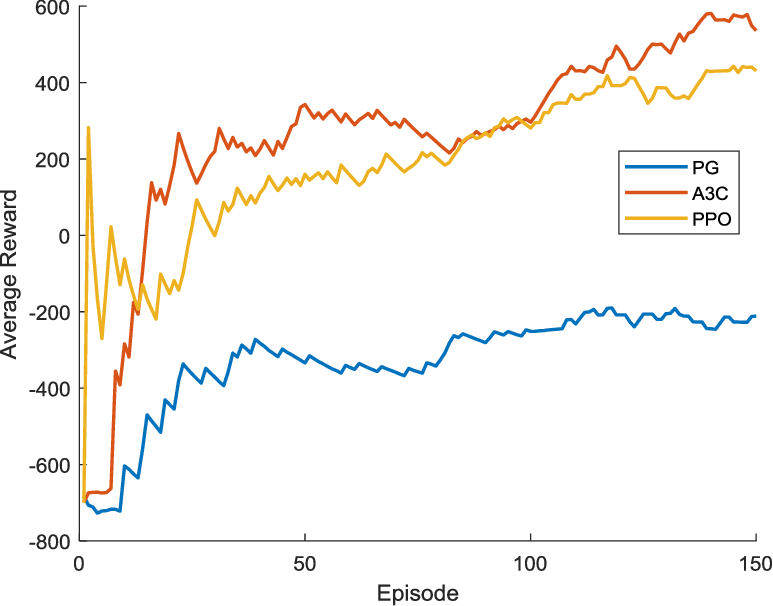}%
\caption{The comparison of average performance in three RL approaches including PG, A3C , and PPO }
\label{Fig07}
\end{figure}

\begin{table} 
\setlength{\tabcolsep}{0.6pt}
\caption{ Comparing the performance of A3C, PPO, PG  {trained with E1 and tested} on E2 and E3 environments}
\label{T1}
\begin{tabular}{|c||c||c||c|}
\hline
Env~/~RL Type&~~A3C~~&~~PPO~~& ~~PG ~~ \\

\hline
 E2&97.9&98.6& 91\\
\hline
E3   &91.6&95.1&89.6) \\
\hline
\end{tabular}

\end{table}

\begin{figure}[t!]
\centering
\includegraphics[width=.7\textwidth]{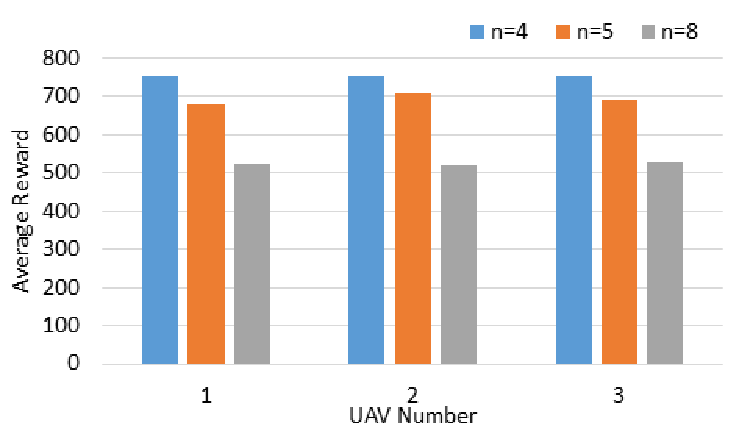}%
\caption{The effect of three different discrete actions (4,5,8)}
\label{Fig08}
\end{figure}

\begin{figure}[t!]
\centering
\includegraphics[width=.7\textwidth]{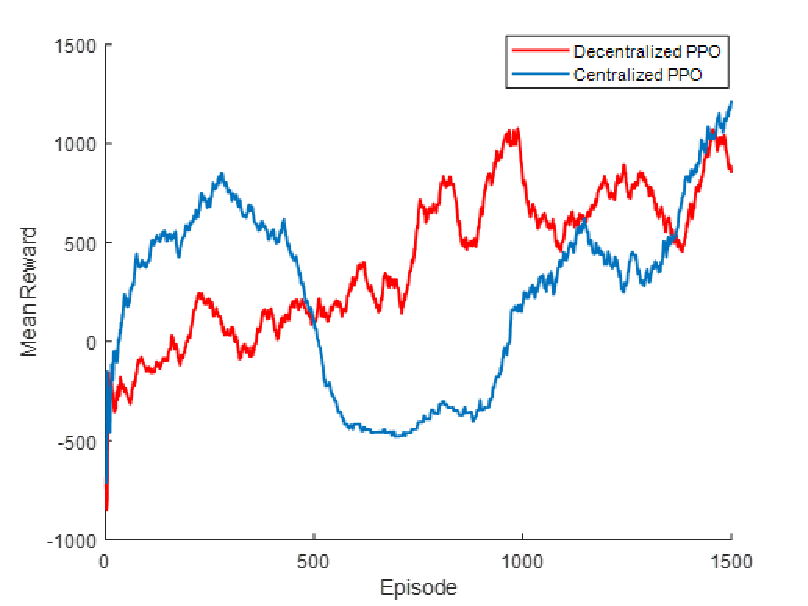}%
\caption{The difference between centralized and decentralized RL training }
\label{Fig09}
\end{figure}
\begin{figure}[t!]
\centering
\includegraphics[width=.7\textwidth]{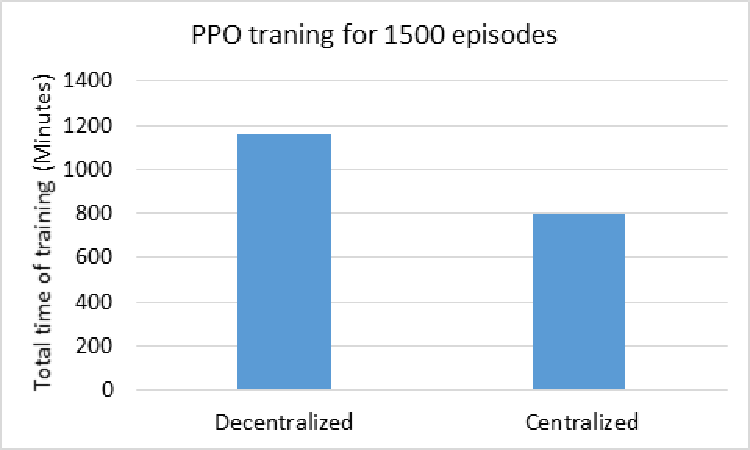}%
\caption{The centralized and decentralized time of training}
\label{Fig10}
\end{figure}

\begin{figure}[t!]
\centering
\includegraphics[width=.7\textwidth]{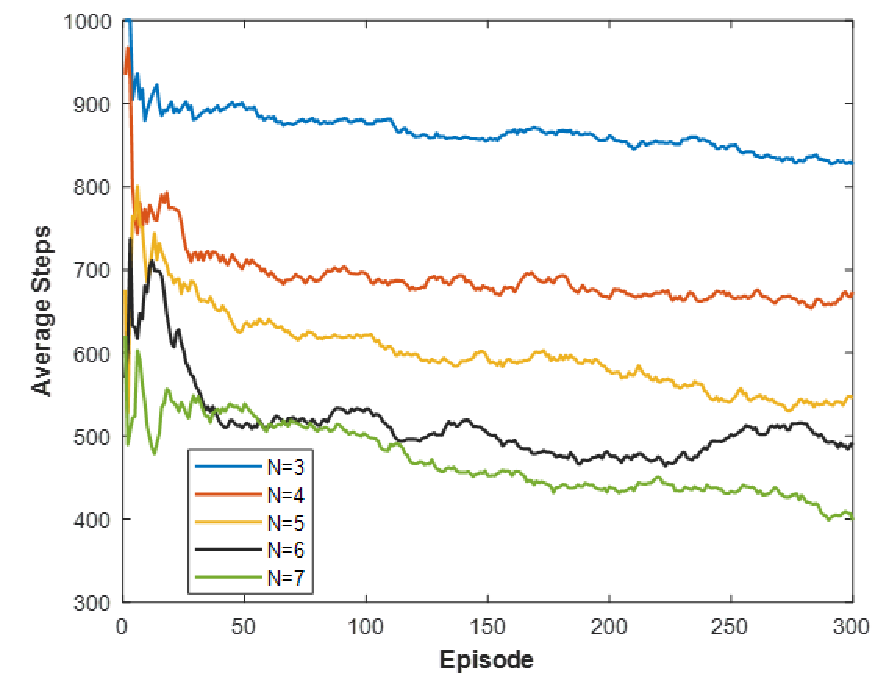}%
\caption{The effect of number of UAVs in number of steps}
\label{Fig11}
\end{figure}

\begin{figure}[t!]
\centering
\includegraphics[width=.7\textwidth]{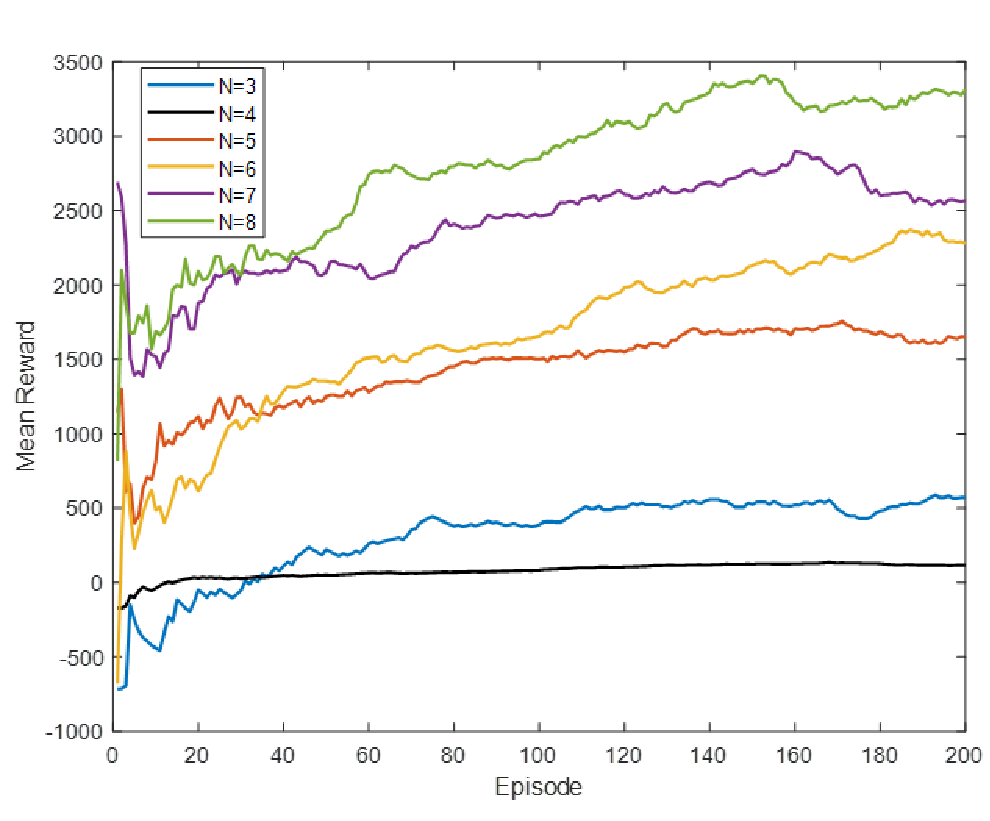}%
\caption{The effect of number of UAVs in average reward}
\label{Fig12}
\end{figure}

\begin{figure}[t!]
\centering
\includegraphics[width=.7\textwidth]{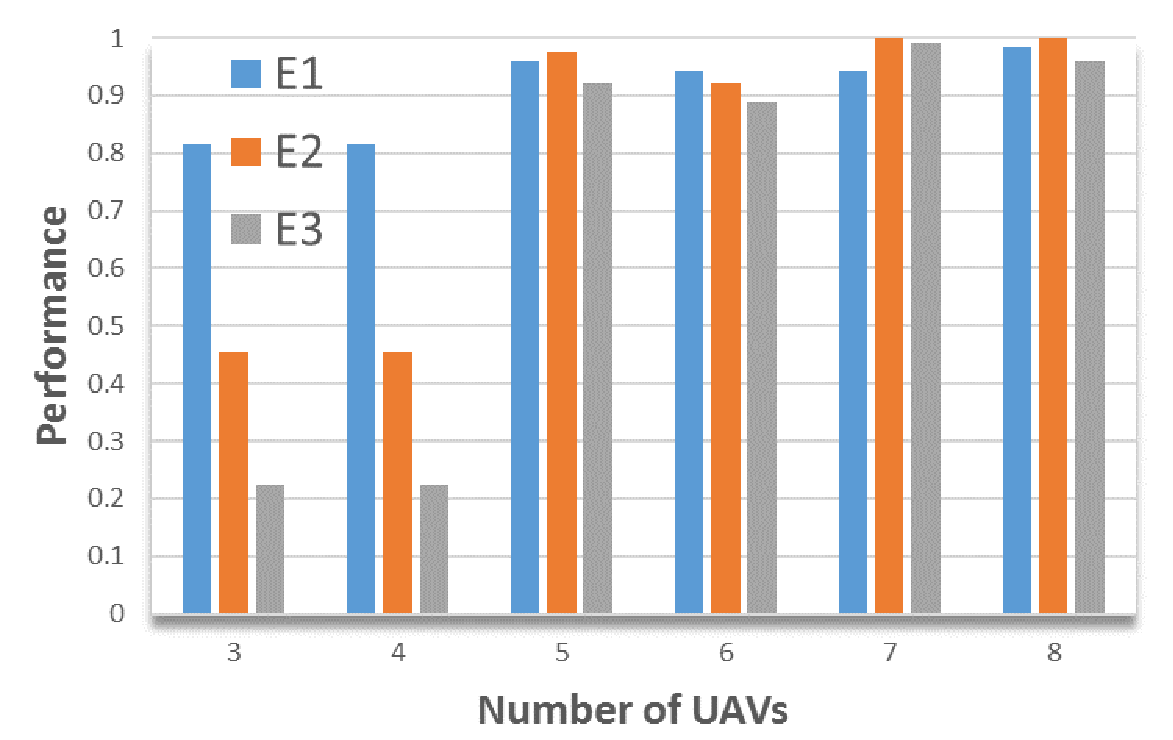}%
\caption{The effect of number of UAVs into the total performance in E1, E2, and E3 environments}
\label{Fig13}
\end{figure}

\begin{figure}[t!]
\centering
\includegraphics[width=0.7\textwidth]{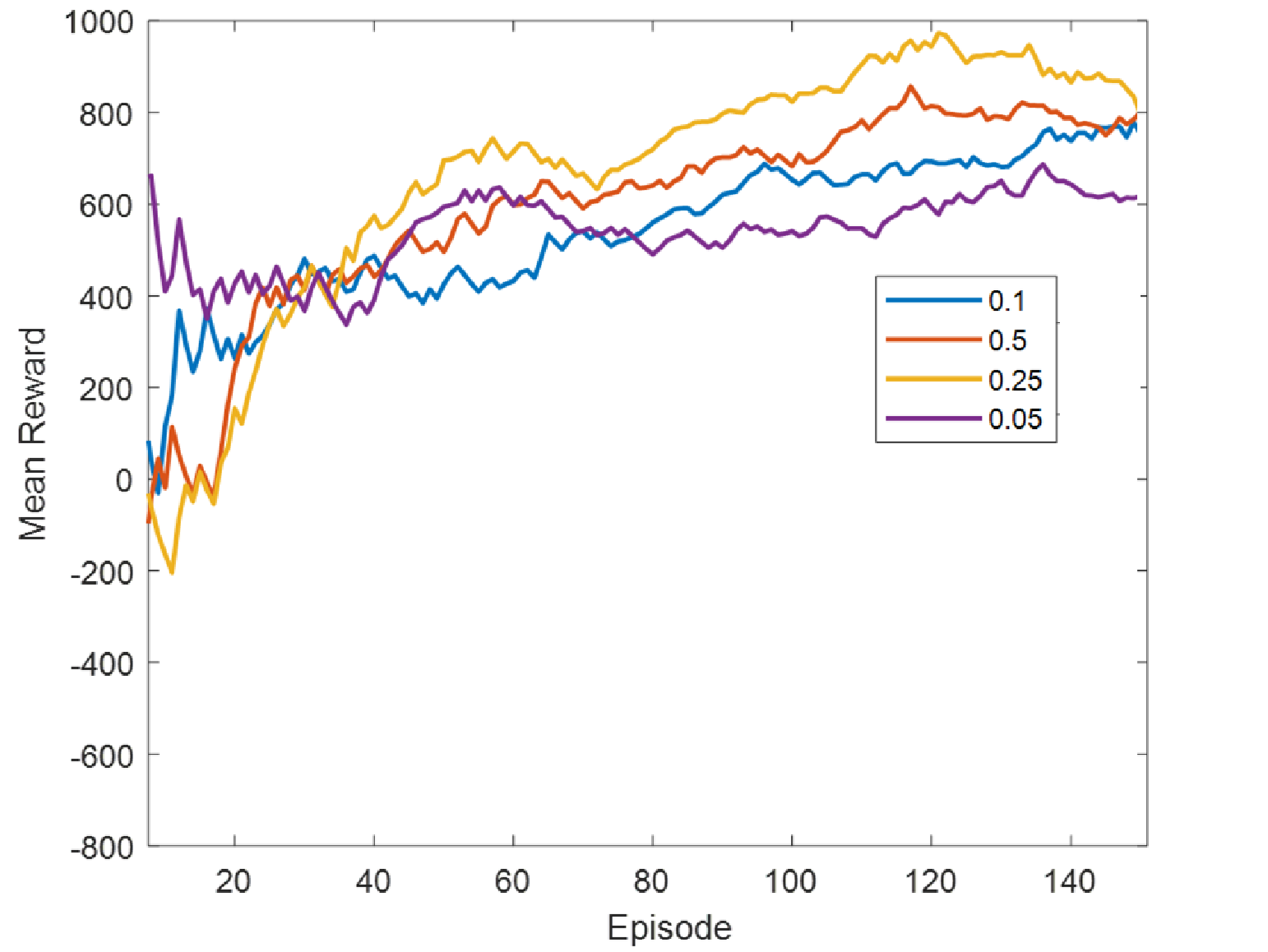}%
\caption{The effect of entropy hyperparameter in the performance }
\label{Fig14}
\end{figure}

\begin{figure}[t!]
\centering
\includegraphics[width=.7\textwidth]{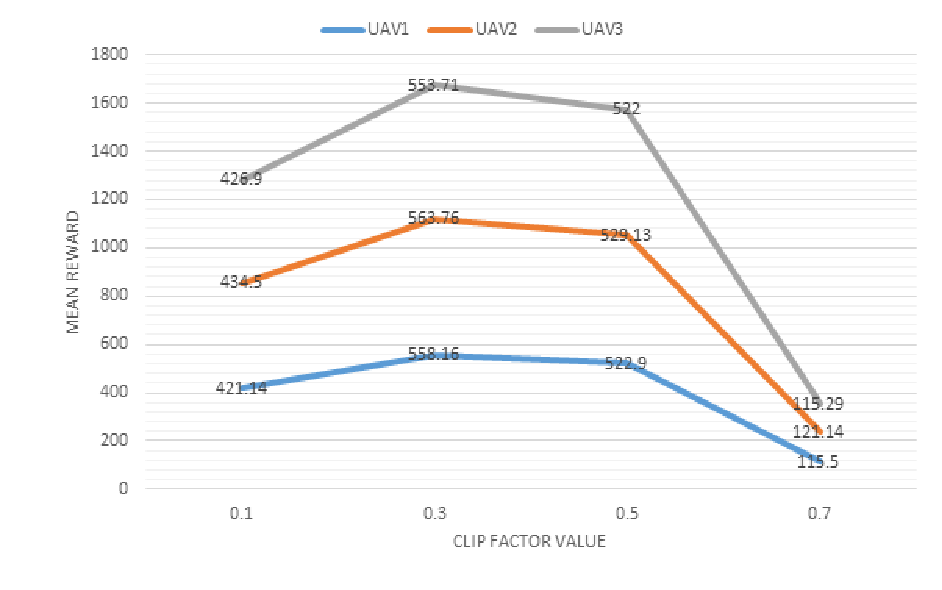}%
\caption{The effect of clip factor hyperparameter in the performance of a mission with 3 UAVs}
\label{Fig15}
\end{figure}

\begin{figure}[t!]
\centering
\includegraphics[width=.7\textwidth]{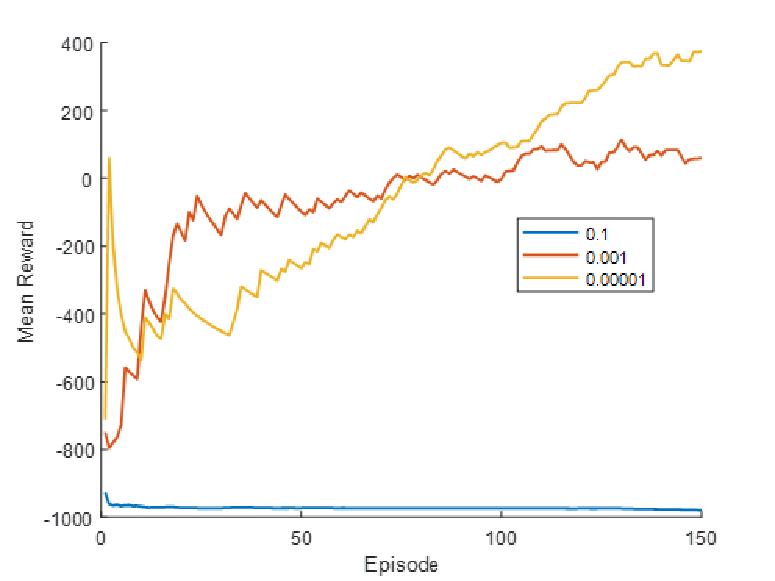}%
\caption{The effect of learning rate hyperparameter in the performance }
\label{Fig16}
\end{figure}

\begin{figure}[t!]
\centering
\includegraphics[width=.7\textwidth]{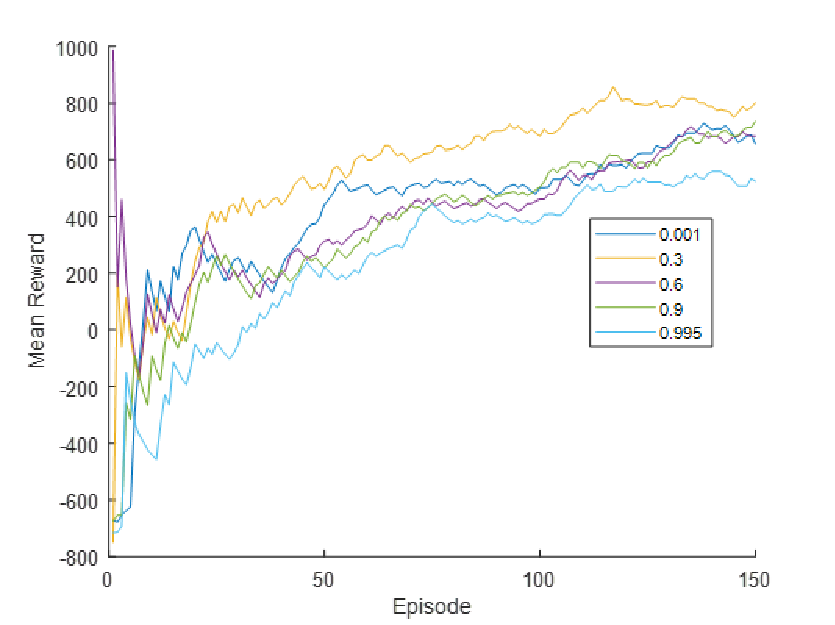}%
\caption{The effect of discount factor hyperparameter in the performance}
\label{Fig17}
\end{figure}

\begin{figure}[t!]
\centering
\includegraphics[width=.7\textwidth]{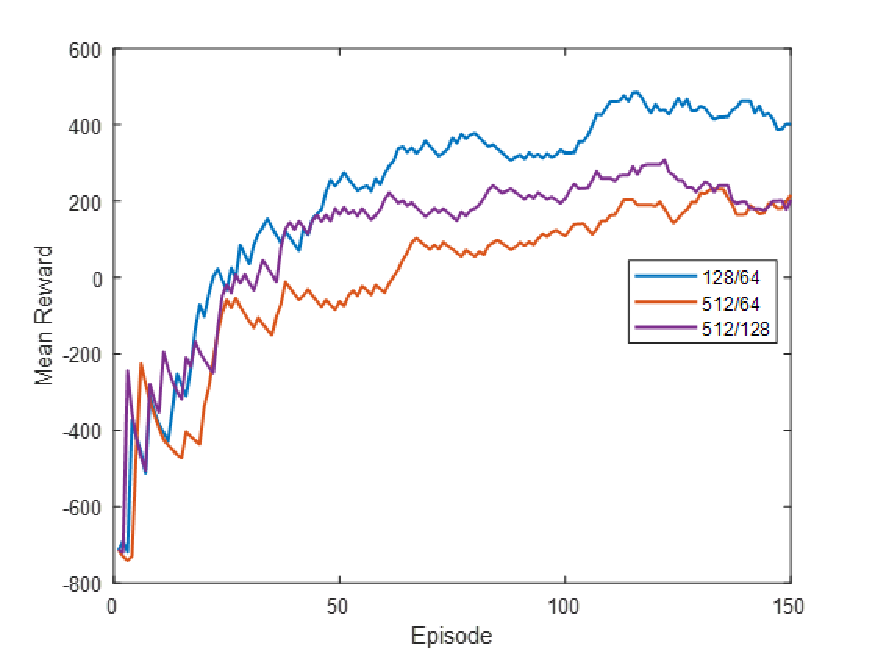}%
\caption{The effect of batch size hyperparameter in the performance}
\label{Fig18}
\end{figure}

\begin{figure}[t!]
\centering
\subcaptionbox {}{ \includegraphics[width=0.65\textwidth]{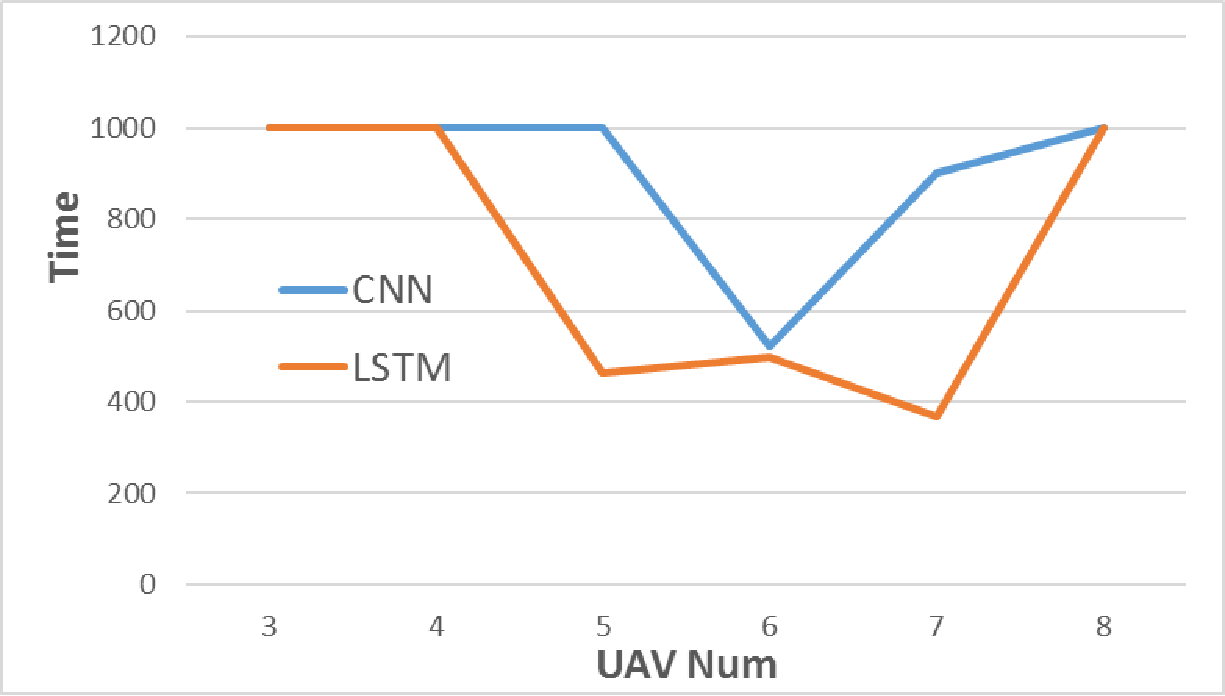}}
\\ 
\subcaptionbox {}{ \includegraphics[width=0.65\textwidth]{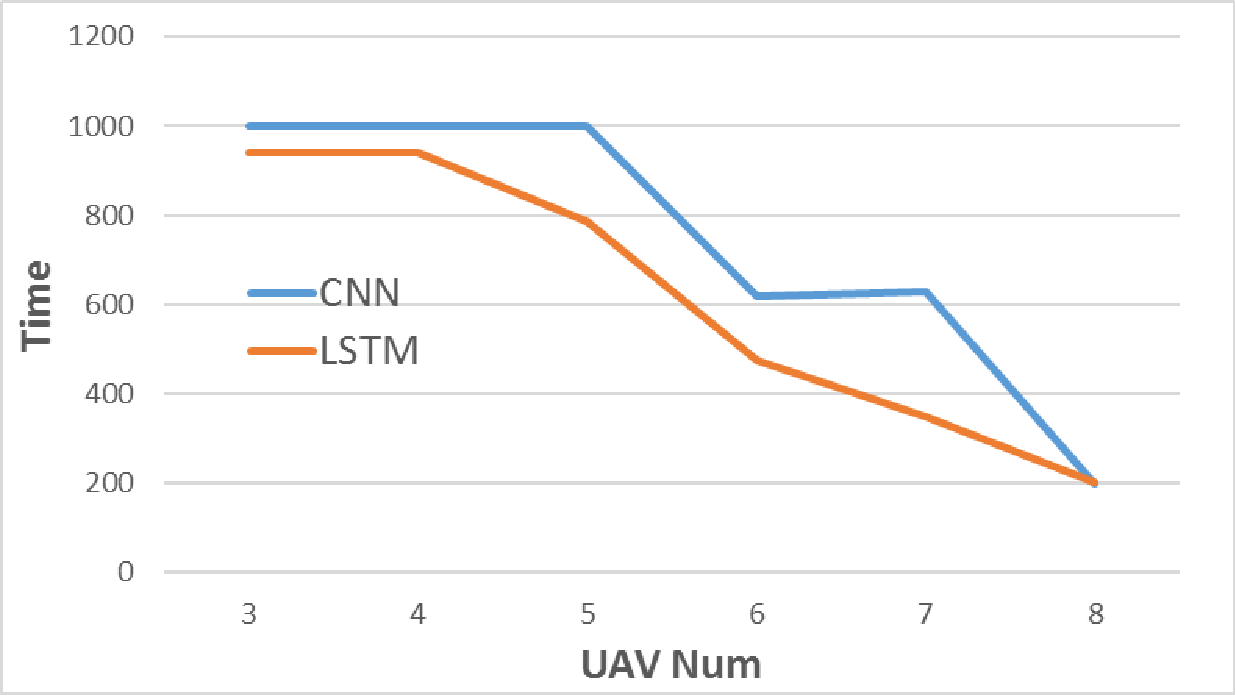}}
\\ 
\subcaptionbox {}{\includegraphics[width=0.65\textwidth]{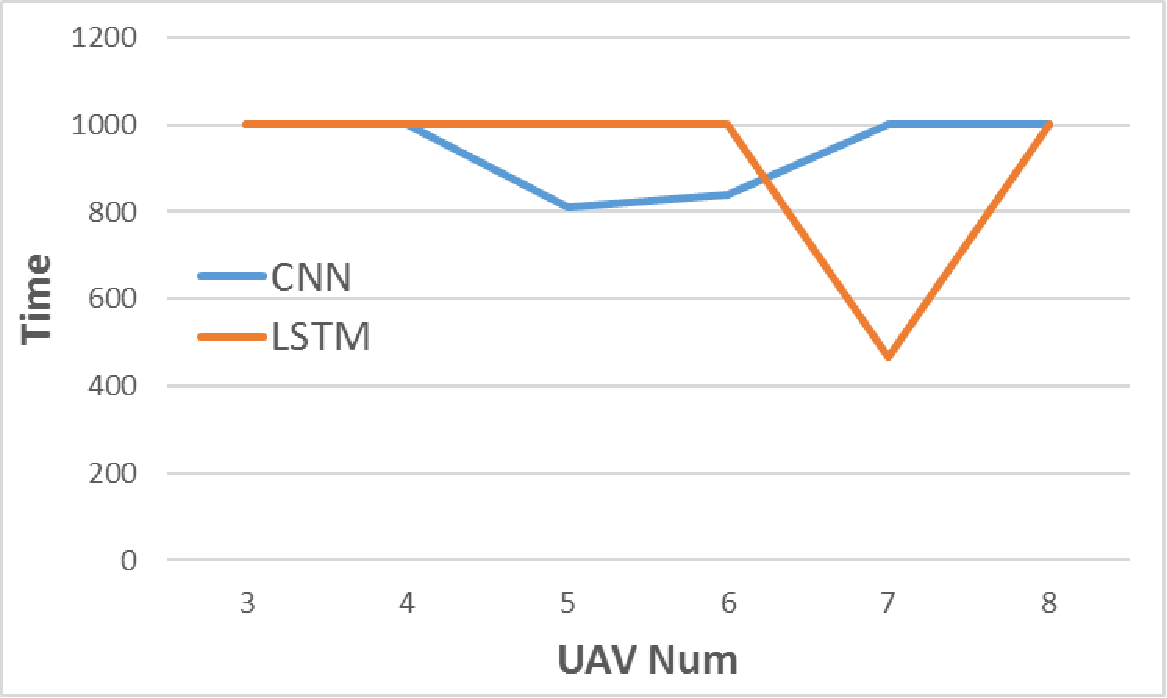}}%
\caption{The coverage time comparison while we use either LSTM or deep CNN in actor in different environments: (a) E1, (b) E2, and (c) E3  }
\label{fig19}
\end{figure}

\begin{figure}[t!]
\centering
\includegraphics[width=0.75\textwidth]{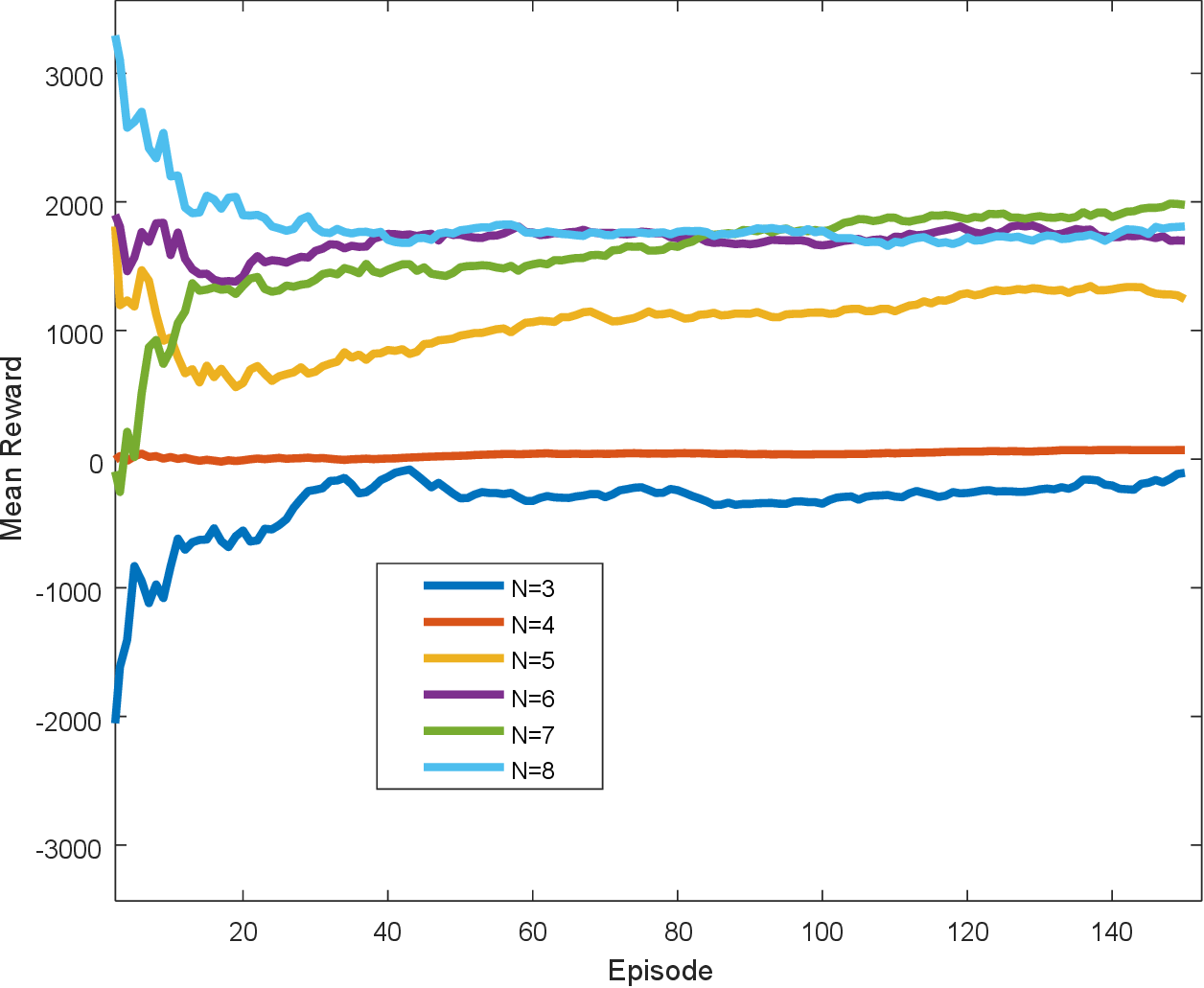}%
\caption{The effect of number of UAVs in average reward using LSTM}
\label{Fig20}
\end{figure}

We compare three on-policy methods in different environments. First, we trained these RL with E1 and then validated the results with E2 and E3. According to Figure 7, the reward of the A3C method is more than that of the PPO method, although the PPO can reach a large reward in a very short time. In this simulation, three UAVs are used for CPP, and the reward graph for all three UAVs is shown in Figure 7. In addition, in a similar experiment, we trained the methods mentioned above (PPO,A3C,PG) with a maximum number of episodes in the E1 environment. Next, Table 3 compares how the trained system might act in environments other than E1 with maximum episodes. The results show the PPO's superiority in terms of total covered area in a limited number of episodes.    

As mentioned earlier, the proposed system has discrete outputs. Therefore, the number of its allowed actions is considered in three states: 4, 5, and 8 allowed actions. After the simulation, it was found that although increasing the allowed actions increases the freedom of decision-making in the UAV,  it reduces the average reward. In addition, 
 adding stops also reduces the rewards compared to 4 actions because if we choose to stop the UAV, it receives a penalty to prevent unnecessary staying unmoved. Figure 8 shows these results.

Although we assume that decision-making is decentralized, our training can be centralized or decentralized. Both methods were considered, and the simulation showed that using the centralized method reduces training speed in the short term, which can be due to the increase in data. Nevertheless, it will have a good reward in the end, as shown in Figures 9 and 10. 

The effect of the number of UAVs on the average steps to complete the search is compared in Figure 11.
The more UAVs there are, the more the search is completed in fewer steps, although sometimes there is little difference in the completion speed according to the environment, such as having 6 and 7 UAVs. Furthermore, The effect of the number of UAVs on the average reward is shown in Figure 12. Usually, the more UAVs there are the more mean reward due to improving the coverage faster.In Figure 13, the exploration performance percentage relative to the number of UAVs is depicted, with a restricted number of iterations. The figure illustrates a decline in performance for E3 and E2 when compared to E1.

\subsection{Hyperparameters}

 We study the effect of different hyperparameter values on the overall performance. Despite its popularity and performance, PPOs also have disadvantages and limitations. One of them is the trade-off between exploration and exploitation. Since PPO uses the same policy for both phases, it may suffer from premature convergence or local optima. To mitigate this problem, PPO relies on entropy adjustment, encouraging the policy to explore more diverse actions. However, this can also reduce the accuracy and stability of the policy. We show how the entropy can change overall performance in Figure 14. 

Another drawback of PPO is sensitivity to the clipping ratio, which determines how much the policy can change between updates. PPO can be too conservative and slow learning if the cutoff ratio is too small. PPO can be too aggressive and make learning unstable if the clipping ratio is too high. The change of this parameter is related to the changes in the amount of reward and the speed of learning, as shown in Figure 15. The learning rate is related to the strength of each gradient descent update step. If training is unstable and the reward does not increase consistently, it should decrease. The learning rate value has been the same for both actor and critic. Figure 16 shows the different learning rates.

In addition, the discount factor determines how much reinforcement learning agents value rewards in the distant future relative to future rewards. This trend is shown in Figure 17. If $\gamma = 0$, the agent will be completely myopic and learn only about actions that produce immediate rewards. If $\gamma = 1$, the agent evaluates his actions based on the total sum of his future rewards. Finally, the experience horizon is the maximum time steps an agent can accumulate during a single experience episode. The user sets this hyperparameter before training the agent, and it depends on the specific task/environment in which the agent operates. If the agent reaches the experience horizon without ending the episode, it resets and starts a new episode.

The time horizon refers to the maximum number of time steps that an agent can perform in one episode of interaction with the environment. This parameter is a familiar concept when dealing with sequential decision problems where an agent must choose actions over a given period to achieve a goal. For example, in a chess game, the time horizon is the maximum number of moves a player can make before the game ends.
In many cases, the time and experience horizons can be set to the same value, especially for simple problems or when the accumulation of experience is relatively fast. However, it may be impossible to set the time and experience horizons to the same value for more complex problems due to computational or memory limitations.
 
Epoch-based training in PPO with minibatch is usually performed by collecting a batch of experience samples by interacting with the environment using the current policy. Then, experience samples are randomly sampled to create a small batch for updating the policy. The effect of different time horizon parameters on the performance is shown in Figure 18.
 

\subsection{Comparison with LSTM}

To evaluate our approach, we trained the PPO in the first environment and tested its performance in other environments. As LSTM has a time series prediction feature, our results,  in Figure 19, show an improvement in overall coverage time for different environments, except for the third environment, which was exceptionally complex. We present the rewards of this network by episodes in Figure 20.

\subsection{The processing time}
{In this subsection, we will be discussing the processing time for the proposed approaches for a fleet of three UAVs in a 12x12 E1 environment. When we use PPO-trained RL, the average processing time for each decision-making step will take approximately 0.001 seconds. However, it takes 1020 minutes of pre-training as displayed in Table 4. To demonstrate the difference between RL and other heuristic options such as Multi-objective evolutionary algorithm based on decomposition (MOEA/D)  and genetic algorithm (GA) \cite{farid2022evolutionary}, we compare the two in Table 5. Please note that MOEA/D is multi-objective while GA is single objective optimizers. In both MOEA/D and GA, the UAVs calculate the tour of waypoints offline, and then follow the waypoints while RL provides an opportunity for the UAVs to start exploration from the initial state and continue until they cover the area of interest. From this perspective, if we apply MOEA/D and GA as a planner, then we require extended time for the simple 144-cell environment comparing with RL. In addition, RL lower amount of computation is vital for embedded processors on board.}

\begin{table} 
\setlength{\tabcolsep}{0.6pt}
\caption{ The processing time of fleet of 3 UAVs for training and execution}
\label{T1}
\begin{tabular}{|c||c|}
\hline
RL execution time (second) & RL training time (minutes) \\

\hline
0.001262 & 1020\\

\hline
\end{tabular}

\end{table}

\begin{table} 
\setlength{\tabcolsep}{0.6pt}
\caption{ The processing time of fleet of 3 UAVs for heuristic MOEA/D and GA}
\label{T1}
\begin{tabular}{|c||c|}
\hline
MOEA/D (seconds) & GA  (seconds) \\

\hline
367.4&4.42	
\\

\hline
\end{tabular}

\end{table}

\section{ Conclusion and Future Work}

{Multi-agent} mapping/exploring unknown areas is a very challenging task. The unknown feature of exploration requires a robust algorithm that can deal with the unknown cases. One of the exciting points of reinforcement learning is that the UAVs can learn about different scenarios that might take place in practice. So, the UAVs can be more reliable using RL when the trained scenarios differ from the specific situation. We used the on-policy feature, which updates the training in each new data using the PPO algorithm. In this paper, we used convolutional neural networks to extract the features in the training phase. Next, after the training finishes, we used decentralized decision-making on each UAV. To test the performance, we showed that the system can be robust in environments that it is not trained for. In addition, using LSTM improves overall performance. 

In the near future, we plan to add more realistic constraints in the simulator, such as weather conditions. In addition, adding heterogeneity in the fleet of robots can be another direction, such as combining fixed-wing, multi-rotor UAVs, and unmanned ground vehicles. We also plan to integrate off-policy learning techniques into PPO to take advantage of instance reuse. Additionally, working on novel reward functions with several non-homogeneous rewards can be an option. {The assumption of a static environment without dynamic obstacles or changes during exploration can be seen as limitation in real-world scenarios. Addressing this limitation could enhance the practical applicability of the proposed approach in the future work}. Furthermore, converting discrete actions into continuous ones can improve performance, especially in complex environments.  Finally, combining real-time reinforcement learning \cite{ramstedt2019real, doukhi2021deep} might improve the RL performance, and we leave it as a future work.

\section*{Data availability}
{Data used in this study are available upon reasonable request.}

\section*{Author Contributions} Methodology, Software, Writing - original draft, Ali Moltajaei Farid;  Writing - review and editing, Jafar Roshanian, Malek Mouhoub.

\subsection*{Ethical and Informed Consent for Data Used}
No ethical data in this paper.

\section*{Declarations}

\subsection*{Conflict of interest}
 {The author declares that there is no conflict of
interest.}

\bibliography{sn-article}

\end{document}